\documentclass[aps,prd,twocolumn,showpacs,preprintnumbers,amsmath,amssymb,superscriptaddress]{revtex4}

\usepackage{color}
\usepackage{graphicx}
\usepackage{dcolumn}
\usepackage{bm}
\usepackage[hypertex]{hyperref}

\newcommand{\bea}{\begin{eqnarray}}
\newcommand{\eea}{\end{eqnarray}}
\newcommand{\pd}{\partial}

\begin{document}

\title{
Momentum space metric,
non-local operator,
and
topological insulators
}

\author{Shunji Matsuura}
\affiliation{Kavli Institute for Theoretical Physics,
 	     University of California, 
 	     Santa Barbara, 
 	     CA 93106, 
 	     USA}

\author{Shinsei Ryu}
\affiliation{
Department of Physics, University of California, Berkeley, CA 94720, USA
            }

\date{\today}

\begin{abstract}

Momentum space of a gapped
quantum system is a metric space: 
it admits a notion of distance 
reflecting properties of its quantum ground state.
By using this quantum metric, we investigate geometric properties of 
momentum space.
In particular, we
introduce a non-local operator which represents distance square in real space and
show that this corresponds to the Laplacian in curved momentum 
space, and also derive its path integral representation in momentum space.
The quantum metric itself measures the second cumulant 
of the position operator
in real space, much like the Berry gauge potential measures the first cumulant or 
the electric polarization in real space. 
By using the non-local operator and the metric, we study some aspects of topological phases such as topological invariants, the cumulants and topological
phase transitions. 
The effect of interactions to the momentum space geometry is also discussed.

\end{abstract}


\maketitle


\section{Introduction}

Topological insulators (superconductors) have been attracting a lot of interest both theoretically
and experimentally.
Similar to trivial insulators, they have a band gap in the bulk.
The striking feature that distinguishes topological insulators (superconductors) 
from trivial ones is that they have gapless modes on the boundaries.
In other words, boundaries of topological insulators are totally metallic.
These gapless modes are robust under perturbations
and cannot be gapped without going through a quantum phase transition,
i.e.,
the gapless modes are topologically protected.
One of the best-known examples of the topological phases is the integer
quantum hall effect (IQHE) in 2+1 dimensions
where the Hall conductance $\sigma_{xy}$ is quantized.


Recent excitement on topological phases comes from the discoveries of new topological phases; 
the quantum spin Hall effect (QSHE) in two space dimensions 
and the $\mathbb{Z}_2$ topological insulator in three space dimensions 
\cite{KaneMele,Roy,moore07,bernevig06,konig07,Fu06_3Da,Qi2008,hasan07}.
The existence of these new phases suggests that the topological phases are relatively generic 
for electron systems: they are not restricted to two space dimensions 
nor they do not necessarily require time reversal symmetry breaking.

Following these discoveries, the exhaustive classification of 
topological insulators and superconductors 
of non-interacting fermionic systems was proposed in
Refs.\ \cite{SRFL,Ryu09,Kitaev}.
The classification is based on discrete symmetries that are generic in quantum systems; 
time reversal symmetry, charge conjugation and chiral (sublattice) symmetry.
All the topological phases are characterized by certain types of K-theory charges; 
$\mathbb{Z}$ Chern/winding numbers and $\mathbb{Z}_2$ numbers.
(For an introduction of K-theory charge for fermi surfaces, see
\cite{Horava:2005jt}.)
These quantities are defined in momentum space:
the Chern number is associated with the Berry curvature (gauge potential),
the winding number is given by a Wess-Zumino-Witten type term constructed from a projection operator,
and $\mathbb{Z}_2$ charge is given by some gauge invariant product of time reversal polarization 
at time reversal invariant points.
The well-known example is again the IQHE: the Hall conductance $\sigma_{xy}$ is given by
the first Chern number of the momentum space Berry curvature via the TKNN formula 
\cite{Thouless82,Kohmoto}.
For time reversal symmetric systems, Berry curvature does not give a non-trivial Chern number.
Instead, the $\mathbb{Z}_2$ number, which basically counts the number of Kramers pairs or Dirac points,
gives the topological charge
\cite{KaneMele,moore07,Roy,Fu06_3Da,FuKane1dz2}.

It is natural to ask if there are other functions or quantities in momentum space that may 
also play a role. 
It is known that the Berry gauge potential is not the only quantity that we can define in momentum space.
For instance, we can generalize the idea of the Berry gauge potential to higher order tensors such as a metric.
Given the fact that topological invariants constructed out of metric play a role in general relativity,
it would be interesting to investigate physical properties of the momentum space metric to see
if it captures new aspects of topological phases.
The momentum space metric (the quantum metric, or the Bures metric) has considered in
Refs.\ \cite{Vanderbilt,Martin} in the context of localization and polarization of Wannier functions and a relation to the cumulants.                                                                                                                                                                                                                                                        
In this paper, we explore a connection 
between the momentum space metric
and physically observable quantities.
Especially we propose a non-local operator that corresponds to 
the Laplacian 
in {\it curved} momentum space.
We will also study some representatives of topological insulators
from the point of view of the momentum space metric.

The organization of this paper is as follows.
In section \ref{review-sec}, we review the construction of the Berry gauge potential and the Bures metric (the quantum metric).
In section \ref{non-local ope}, we first review the known properties of the Bures metric
and then define and investigate a non-local operator corresponding to the Laplacian in curved
momentum space.
In section \ref{case-studies}, we study several topological phases by using the Bures metric,
especially for symmetry classes A (unitary) and AII (symplectic, spin-orbit).

\section{Momentum space gauge field and gravity} 
\label{review-sec}
\subsection{setup} 

In this section, we will explore momentum space geometry 
and topology 
by studying gauge field (the Berry gauge field)
and metric (the Bures metric).
Consider a tight-binding Hamiltonian,
\begin{eqnarray}
H 
\!\!&=&\!\!
\sum_{\mathbf{r},\mathbf{r}'} 
\psi^{\dag}(\mathbf{r})\, 
\mathcal{H}(\mathbf{r},\mathbf{r}')\, 
\psi(\mathbf{r}'), 
\end{eqnarray}
where 
$\psi(\mathbf{r})
$ 
is an $N_f$-component 
fermion annihilation operator,
and index $\mathbf{r}$ labels a site
on a $d$-dimensional lattice
(the internal indices are suppressed).
Each block in the single particle Hamiltonian 
$\mathcal{H}(\mathbf{r},\mathbf{r}')$ is 
an $N_f\times N_f$ matrix,
satisfying the hermiticity condition
$\mathcal{H}^{\dag}(\mathbf{r}',\mathbf{r}) 
=\mathcal{H}(\mathbf{r},\mathbf{r}')$. 
and we assume the total size of the 
single particle Hamiltonian is
$N_f V\times N_f V$,
where $V$ is the total number of lattice sites. 
The components in $\psi(\mathbf{r})$
can describe, e.g., 
orbitals or spin degrees of freedom,
as well as
different sites within a crystal unit cell
centered at $\mathbf{r}$.

Provided the system has 
translational symmetry,
\begin{eqnarray}
\mathcal{H}(\mathbf{r},\mathbf{r}') = 
\mathcal{H}(\mathbf{r}-\mathbf{r}'),
\end{eqnarray}
with periodic boundary conditions in each spatial direction
(i.e., the system is defined on a torus $T^d$),
we can perform the Fourier transformation and obtain 
in momentum space
\begin{eqnarray}
H 
\!\!&=&\!\!
\sum_{\mathbf{k} \in \mathrm{BZ}} 
\psi^{\dag}(\mathbf{k}) \,
\mathcal{H}(\mathbf{k})\, 
\psi (\mathbf{k}) ,
\end{eqnarray}
where the crystal momentum $\mathbf{k}$
runs over the first Brillouin zone (BZ), 
and the Fourier component of the fermion operator
and the Hamiltonian are given by 
\begin{eqnarray}
\psi(\mathbf{r})
=
V^{\frac{-1}{2}}
\sum_{\mathbf{k}\in \mathrm{BZ}}
e^{{i} \mathbf{k}\cdot \mathbf{r}}
\psi(\mathbf{k}),
\,\,
\mathcal{H}(\mathbf{k})
=
\sum_{\mathbf{r}} 
e^{- {i} \mathbf{k}\cdot \mathbf{r}}
\mathcal{H}(\mathbf{r}),
\end{eqnarray}
where 
$V$ is the total number of sites, 
and 
$\mathcal{H}(\mathbf{k})$ is an $N_{f} \times N_{f}$
matrix.

The Bloch Hamiltonian $\mathcal{H}(\mathbf{k})$
is diagonalized by 
\begin{eqnarray}
\mathcal{H}(\mathbf{k})
|u^{{a}}(\mathbf{k}) \rangle
=
E^{a} (\mathbf{k})|u^{a}(\mathbf{k}) \rangle,
\quad
{a} = 1,\ldots, N_{f},
\end{eqnarray}
where
and $|u^{{a}}(\mathbf{k}) \rangle$ is the 
${a}$-th Bloch
wavefunction with energy 
$E^{{a}} (\mathbf{k})$.
The fermion operator can be expanded in terms of 
the eigen functions as
\begin{eqnarray}
\psi_i (\mathbf{k})
=
u_{i}{ }^{{a}}(\mathbf{k}) 
\chi_{{a}}(\mathbf{k}),
\quad 
i, {a} = 1,\ldots, N_{f},
\end{eqnarray}
where 
$\chi_{{a}}(\mathbf{k})$ represents 
a fermionic operator in the eigen basis
and given by
\begin{eqnarray}
\chi_{{b}}(\mathbf{k})
=
\big[u_{i}{ }^{b}(\mathbf{k})
\big]^* 
 \psi_i (\mathbf{k}).
\end{eqnarray}

We assume that
there is a finite gap at the Fermi level, 
and therefore
we obtain a unique ground state 
$|\Phi (\mathbf{k})\rangle$
for each $\mathbf{k}$
by filling all states below
the Fermi level. 
(In this paper, we always adjust 
$E^{\hat a}(\mathbf{k})$ in such a way 
that the Fermi level is at zero energy.) 
We assume there are $N_-$ ($N_+$)
occupied (unoccupied) Bloch wavefunctions 
for each $\mathbf{k}$
with $N_{+} + N_{-} = N_{f}$.
We call the set of filled Bloch wavefunctions 
$
\{
|u^{-}_{\hat a}(\mathbf{k}) \rangle
\}
$,
where 
hatted indices $\hat{a}=1,\ldots, N_-$ 
labels the occupied bands only.
The many-body ground state $|\Phi \rangle$ is the filled Fermi sea
\begin{eqnarray}
|\Phi \rangle
=
\bigotimes_{\mathbf{k}\in \mathrm{BZ}}
|\Phi (\mathbf{k})\rangle, 
\quad
|\Phi (\mathbf{k})\rangle
=
\prod^{N_-}_{\hat{a}=1}
\chi^{\dag}_{\hat{a}}(\mathbf{k})|0\rangle.
\label{gs for non-interacting}
\end{eqnarray}

\subsection{Berry connection
and Bures metric (quantum metric)}

The momentum space gauge field and metric can be introduced 
by comparing two states 
$|\Phi(\mathbf{k}_1)\rangle$
and
$|\Phi(\mathbf{k}_2)\rangle$
at different momenta $\mathbf{k}_{1,2}$. 
From the overlap 
$
\langle\Phi(\mathbf{k}) | \Phi(\mathbf{k}+d\mathbf{k})\rangle
$,
we define the Berry connection \cite{Berry, WilcZee} 
$\mathcal{A}(\mathbf{k})=A_{\mu}(\mathbf{k})dk_{\mu}$ 
by the expansion
\begin{eqnarray}
\langle
\Phi(\mathbf{k}) | \Phi(\mathbf{k}+d\mathbf{k})
\rangle
=
1 + A_{\mu}(\mathbf{k}) dk_{\mu} + \cdots, 
\end{eqnarray}
where
$\mu=1,\cdots,d$.
Similarly,
to define the momentum space metric,
we consider the quantum distance 
$D_{\mathbf{k}_1,\mathbf{k}_2}$ between
two states
$|\Phi(\mathbf{k}_1)\rangle$
and
$|\Phi(\mathbf{k}_2)\rangle$
by
$
D^2_{\mathbf{k}_1,\mathbf{k}_2} = 
1- 
|
\langle\Phi(\mathbf{k}_1)| \Phi(\mathbf{k}_2)\rangle
|^2 
$
~\cite{Provost:1980nc}.
When these two states are infinitesimally close to each other, 
$\mathbf{k}_1=\mathbf{k}$,
$\mathbf{k}_2=\mathbf{k} + d\mathbf{k}$, 
the quantum distance can be expanded as 
\begin{eqnarray}
|
\langle\Phi(\mathbf{k})| \Phi(\mathbf{k}+ d\mathbf{k})\rangle
|^2 
= 
1-
g_{\mu\nu}(\mathbf{k})
dk_{\mu}
dk_{\nu}
+
\cdots. 
\end{eqnarray}
This defines the quantum metric $g_{\mu\nu}(\mathbf{k})$.

For the case when 
$|\Phi(\mathbf{k})\rangle$
is given by a Slater determinant, 
the overlap 
$\langle\Phi(\mathbf{k}_1) | \Phi(\mathbf{k}_2)\rangle$
is computed as a determinant of 
an $N_-\times N_-$ matrix
made of Bloch states, 
\begin{eqnarray}
\langle\Phi(\mathbf{k}_1) | \Phi(\mathbf{k}_2)\rangle
\!\!&=&\!\!
\mathrm{det}\,
S(\mathbf{k}_1,\mathbf{k}_2), 
\nonumber \\
S_{\hat{a}\hat{b}}(\mathbf{k}_1,\mathbf{k}_2)
\!\!&:=&\!\!
\langle u^-_{\hat{a}}(\mathbf{k}_1) | u^-_{\hat{b}}(\mathbf{k}_2)\rangle, 
\end{eqnarray}
where 
$| u^-_{\hat{a}}(\mathbf{k})\rangle
$
($\hat{a}=1,\cdots, N_-$),
are the Bloch wavefunction for
filled bands. 
Hence the Berry connection 
and the Bures metric are given by 
\begin{eqnarray}
A_{\mu}(\mathbf{k})
\!\!&=&\!\!
\langle u^{-}_{\hat a}(\mathbf{k}) | 
\partial_{\mu} u^{-}_{\hat a}(\mathbf{k}) \rangle, 
\\
g_{\mu\nu}(\mathbf{k})
\!\!&=&\!\!
\frac{1}{2}
\Big[
\langle \partial_{\mu} u^-_{\hat{a}}| \partial_{\nu} u^-_{\hat{a}} \rangle 
+
\langle \partial_{\nu} u^-_{\hat{a}}| \partial_{\mu} u^-_{\hat{a}} \rangle 
\nonumber \\
&&
\qquad 
+
2
\langle u^-_{\hat{a}}| \partial_{\mu} u^-_{\hat{b}} \rangle 
\langle u^-_{\hat{b}}| \partial_{\nu} u^-_{\hat{a}} \rangle
\Big]
\\
\!\!&=&\!\!
\mathrm{Re}\,\langle \partial_{\mu} u^-_{\hat{a}}| \partial_{\nu} u^-_{\hat{a}} \rangle 
-
\langle \partial_{\mu}  u^-_{\hat{a}}| u^-_{\hat{b}} \rangle 
\langle u^-_{\hat{b}}| \partial_{\nu} u^-_{\hat{a}} \rangle, 
\nonumber
\end{eqnarray}
where 
repeated indices are summed, and 
$\partial_{\mu}=\partial_{k_{\mu}}$, etc.

For the case with more than one occupied bands, 
one can also introduce the non-Abelian Berry connection by 
\begin{eqnarray}
&&
\mathcal{A}^{\hat a \hat b}(\mathbf{k})
=
A^{\hat a \hat b}_{\mu}(\mathbf{k}) {d} k_{\mu} 
=
\langle u^{-}_{\hat a}(\mathbf{k}) | d u^{-}_{\hat b}(\mathbf{k}) \rangle,
\end{eqnarray}
where
$A^{\hat a \hat b}_{\mu}
=
-(A^{\hat b \hat a}_{\mu})^*$. 
A non-Abelian generalization of the metric 
is discussed in Refs.\ \cite{MaChenFanLiu}.

The above formula for the 
gauge field and metric can compactly summarized 
in terms the spectral projector 
onto the filled Bloch states
and the ``$Q$-matrix'', which are defined by
\begin{eqnarray}
P(\mathbf{k})
:=
\sum_{\hat{a}}
|u^{-}_{\hat a}(\mathbf{k}) \rangle
\langle u^{-}_{\hat a}(\mathbf{k}) |, 
\, \,
Q(\mathbf{k}) := 1 - 2P(\mathbf{k}).
\label{P and Q}
\end{eqnarray}
Then, the U(1) part of the Berry curvature
the metric are given by 
\begin{eqnarray}
F_{\mu\nu}(\mathbf{k})
\!\!&=&\!\!
F^{\hat{a} \hat{a}}_{\mu\nu}(\mathbf{k})
=
\frac{1}{4}
\mathrm{tr}\, \left[
Q \partial_{\nu} Q \partial_{\mu} Q
\right], 
\nonumber \\
g_{\mu\nu}(\mathbf{k})
\!\!&=&\!\!
\frac{1}{8} 
\mathrm{tr}\, \left[
\partial_{\mu} Q \partial_{\nu} Q
\right]. 
\label{berry-metric-P-Q}
\end{eqnarray}
I.e., 
$g_{\mu\nu}$ and
$F_{\mu\nu}$ are the symmetric and
antisymmetric part of
$\mathrm{tr}\, \left[ P \partial_{\nu}P \partial_{\mu}P \right]$,
respectively,
$2 g_{\mu\nu} + F_{\mu\nu}
=
- 
\mathrm{tr}\, \left[ P \partial_{\nu}P \partial_{\mu}P \right]
$.

\section{Non-local operators 
and moments of the position operator}
\label{non-local ope}

In this section,
we discuss 
the expectation values of the position operator,
and its second cumulant in a quantum ground state. 
For band insulators,
the Berry connection and the Bures metric 
naturally enter in evaluating these expectation values.
This is to some extent expected,
since 
the definition of the Berry connection and the Bures metric  
involves a derivative with respect to momentum, 
this in turn suggests
that it is related to the position operator in real space.
We will first review these ideas developed 
in the theory of macroscopic polarization,
and 
in the context of maximally localized Wannier functions
in solids. 
We will then introduce yet another formalism for the 
second cumulant of the position operator, which is for
the geometric, rather than arithmetic, mean of the 
second cumulant. 
We will show how it is related to the momentum 
space spectral geometry.

\subsection{Marzari-Vanderbilt theory}


Let us consider the following non-local operator: 
\begin{eqnarray}
z_{\boldsymbol{\alpha} }= 
\exp
\left[
\sum_{\mathbf{r}}
{i} \boldsymbol{\alpha} 
\cdot 
\mathbf{r}\rho(\mathbf{r})
\right]
\end{eqnarray}
where $\mathbf{r}=(x_1,x_2,\cdots, x_{d})$ labels 
sites on
the $d$-dimensional hyper cubic lattice
with $N^d$ ($=V$) sites, 
and can be written, in terms of 
the lattice constant $\mathfrak{a}$ as 
\begin{eqnarray}
x_{\mu} = n_{\mu}\times 
\mathfrak{a},
\quad
n_{\mu} = 1,\cdots, N,
\quad
L= \mathfrak{a}N. 
\end{eqnarray}
$\rho(\mathbf{r})$ is the density operator 
at $\mathbf{r}$, 
\begin{eqnarray}
\rho(\mathbf{r}) 
\!\!&=&\!\!
\sum_{i=1}^{N_f}
\psi^{\dag}_i (\mathbf{r})
\psi^{\ }_i (\mathbf{r}). 
\end{eqnarray}

We now choose 
\begin{eqnarray}
\boldsymbol{\alpha} 
= 
2\pi
\hat{\mathbf{n}}_{\mu}/L,
\end{eqnarray}
where $\hat{\mathbf{n}}_{\mu}$
is 
the unit vector in 
the $\mu$-direction 
and consider the expectation value of the non-local operator,
\begin{eqnarray}
\mathfrak{z}_{\mu} 
:= 
\left\langle \Psi \right| 
z_{2\pi \hat{\mathbf{n}}_{\mu}/L}
\left| \Psi\right\rangle. 
\end{eqnarray}
The logarithm of $\mathfrak{z}$ 
defines the expectation of 
the position operator 
in periodic systems
\begin{eqnarray}
\langle
x_{\mu}
\rangle
=
\frac{L}{2\pi}
\mathrm{Im}
\ln 
\mathfrak{z}_{\mu}. 
\end{eqnarray}
(The ``single point formula'' for the macroscopic 
polarization by Resta \cite{Resta94,Resta98, Resta01}.)

For the case of Slater determinant,
this can be evaluated as 
\begin{eqnarray}
\langle
x_{\mu}
\rangle
\!\!&=&\!\!
\frac{L}{2\pi}
\sum_{\mathbf{k}}
\mathrm{tr}\, \ln 
S(\mathbf{k}, \mathbf{k}+\hat{\mathbf{k}}_{\mu}).
\end{eqnarray}
where $\hat{\mathbf{k}}_{\mu}=2\pi
\hat{\mathbf{n}}_{\mu}/L$ and the trace is taken over occupied states. 
In the continuum limit, 
this reduces to the Wilson loop of the 
U(1) part of the Berry connection,
$\langle
x_{\mu}
\rangle
\sim 
\ln 
\exp\oint_{\mathrm{BZ}} \mathrm{tr}\,  \mathcal{A}(\mathbf{k}) 
$.

In general, we can consider the generating function
\begin{eqnarray}
C(\boldsymbol{\alpha})
\!\!&=&\!\!
\left\langle \Psi \right| 
z_{\boldsymbol{\alpha}}
\left| \Psi\right\rangle.
\label{gene-fucn}
\end{eqnarray}
For slater determinants,
this can be evaluated as
\begin{eqnarray}
\ln C(\boldsymbol{\alpha})
=
\sum_{\mathbf{k}}
\mathrm{tr}\, 
\ln 
S(\mathbf{k}, \mathbf{k}+\boldsymbol{\alpha}).
\end{eqnarray}
Taking the derivative with respect to
$\boldsymbol{\alpha}$, 
we generate all cumulants
of the position operators, 
\begin{eqnarray}
&&
\langle x_1^{n_1}x_2^{n_2}\cdots x_d^{n_d} \rangle_c
\nonumber \\
&&
\quad
\simeq 
i^{n}
\left({\pd^{n} \over \pd \alpha_1^{n_1}\pd \alpha_2^{n_2}\cdots \pd \alpha_d^{n_d}}\right)
\ln C(\boldsymbol{\alpha})\Big|_{\alpha=0}.
\end{eqnarray}
where
the derivative with respect to
$\alpha_{\mu}$ should be properly discretized
\cite{Resta01}. 
E.g.,
\begin{eqnarray}
\langle
x_{\mu} x_{\mu}
\rangle_c
=
\left(
\frac{\mathfrak{a}}{2\pi}
\right)^2
\frac{-2}{N_f L}
\ln
\prod_{\mathbf{k}}
\mathrm{det}\,
\big|
S(
\mathbf{k},
\mathbf{k}+
\hat{\mathbf{k}}_{\mu}
)
\big|^{2},
\end{eqnarray}
where $\mu$ is not summed. 
In the continuum, 
the second cumulant is given by
\begin{eqnarray}
\langle
x_{\mu} x_{\nu}
\rangle_c
\!\!&=&\!\!
\frac{v}{N_f (2\pi)^d }\int_{\text{BZ}} d \mathbf{k}\, 
g_{\mu\nu}( \mathbf{k})
=:
\frac{v}{N_f} \Omega_{I,\mu\nu}.
\end{eqnarray}
with $v$ the cell volume.


This expression 
for the second cumulant
should be compared with
$\Omega
= \sum_{\hat a} \left[ 
\langle r^2 \rangle_{\hat a} -\langle {\bf r} \rangle^2_{\hat a}
\right],
$
where 
$
\langle r^2 \rangle_{\hat a}
=
[v/(2\pi)^d]
\langle \partial_{\mu}   u_{\hat a}(\mathbf{k})| 
\partial_{\mu}
u_{\hat a}(\mathbf{k}) \rangle
$
and
$
\langle {\bf r} \rangle_{\hat a}
=
i 
[v/(2\pi)^d]
\int d{\bf k}
\langle u_{\hat a}(\mathbf{k})| \nabla_{{\bf k}} | u_{\hat a}(\mathbf{k}) \rangle
$, 
which was also used in \cite{Vanderbilt}. In that paper, it was showed that in a band model,
$\Omega$ is not gauge invariant and hence cannot directly be related to an observable,
while $\Omega_{I}$ is gauge invariant.
This gauge dependence argument would not be relevant in this paper.
We mostly consider general continuum models.


\subsection{non-local operator}
\label{non-local operator}

We introduce yet another non-local operator $\eta$ by 
\begin{eqnarray}
\eta
\!\!&=&\!\!
\exp
\left\{
\sum_{\mathbf{r}}
\Theta(\mathbf{r})
\left[
\rho(\mathbf{r})
-\bar{\rho}
\right] 
\right\},
\label{eq: def non-local op}
\end{eqnarray}
where $\bar{\rho}$
is the average density,
and 
$\Theta(\mathbf{r})$ is assumed to be written in terms of 
a logarithm of some function, 
\begin{eqnarray}
\Theta(\mathbf{r})
\!\!&:=&\!\!
\ln
\theta(\mathbf{r}).
\end{eqnarray}
For our purpose,
we will choose
\begin{eqnarray}
&&
\theta(\mathbf{r}) 
=
\left[
-2 
\sum^d_{\mu=1} 
\cos \frac{2\pi n_{\mu}}{N} 
+2d 
\right]^{\alpha}
\nonumber \\
&&
\mbox{with}
\qquad 
\alpha=1/N_e
= 
1/(fV)
\end{eqnarray}
where $N_e$ is the total number of electrons,
and $f=N_e/V$ is the filling fraction. 
By expanding the cosine
in the decompactifying limit,
$x_{\mu}/L\to 0$,
\begin{eqnarray}
\ln \theta(\mathbf{r})
\sim 
\ln
\sum^d_{\mu=1}
\left(
\frac{2\pi x_{\mu}}{\mathfrak{a} N} 
\right)^2
=
\ln
\frac{r^2}{\mathfrak{a}^2}
-
\ln
\left(
\frac{2\pi}{N} 
\right)^2, 
\end{eqnarray}
and noting
\begin{eqnarray}
&&
\int d^dr\,
\left[
\ln
\frac{r^2}{\mathfrak{a}^2}
-
\ln
\left(
\frac{2\pi }{N} 
\right)^2
\right]
\left[
\rho(\mathbf{r})-\bar{\rho}
\right]
\nonumber \\
\!\!&=&\!\!
\int d^dr\,
\ln
(r^2/\mathfrak{a}^2)
\left[
\rho(\mathbf{r})-\bar{\rho}
\right],
\end{eqnarray}
where we have used
the charge conservation
$\int d^dr 
[\rho(\mathbf{r})-\bar{\rho}]
=0$,
$\eta$ may be interpreted,
in the continuum limit, as
\begin{eqnarray}
\eta
\!\!&=&\!\!
\exp
\left\{
\alpha 
\int \frac{d^d r}{\mathfrak{a}^d}
\left(\ln r^2/\mathfrak{a}^2 \right) 
\left[
\rho(\mathbf{r})-\bar{\rho}
\right]
\right\}.
\end{eqnarray}
In the first quantization, 
the density operator is given by 
\begin{eqnarray}
\rho(\mathbf{r})
=
\sum_{i=1}^{N_e}\delta^{(d)} (\mathbf{r}-\mathbf{r}_i)
\end{eqnarray}
where $\mathbf{r}_i$ is the position operator 
of the $i$-th particle.
Then, 
\begin{eqnarray}
\exp \alpha \int d^d r\, 
(\ln\, r^2) \rho(\mathbf{r})
=
e^{ \alpha \sum^{N_e}_{i=1} \ln\,  r^2_i  }
= 
\prod^{N_e}_{i=1} 
\left(r^2_i\right)^{\alpha}. 
\end{eqnarray}
Thus, the expectation value
of $\eta$ is nothing but the geometric mean of 
the position operator over $N_e$
fermions. 

We note that 
by choosing $\Theta(\mathbf{r})$
in Eq.\ (\ref{eq: def non-local op})
properly, 
$\eta$ can represent 
several different non-local operators familiar 
in condensed matter context.
When we choose 
\begin{eqnarray}
\theta(\mathbf{r})
= 
e^{{i} \frac{2\pi n_x}{N}},
\quad
\Theta(\mathbf{r})
=
{i} \frac{2\pi n_x}{N},
\end{eqnarray}
$\eta$ is the 
twist operator 
$\mathfrak{z}$
used in the theory of 
electric polarization.
In two dimensions, and when we choose
\begin{eqnarray}
\theta(\mathbf{r})=( z-z_0)^m ,
\quad 
\Theta(\mathbf{r})
= 
\ln (z-z_0)^m ,
\end{eqnarray}
where $z=x+{i}y$,
and $z_0$ is some reference point,
then
$\eta$ is the disorder operator 
studied by Shindou et al. 
\cite{Shindou05, LeeKivelson}.
Yet another non-local operator
in the form (\ref{eq: def non-local op})
is discussed in 
Ref.\ \cite{Ryuentro}
in the context of the entanglement entropy.

\subsection{the non-local operator 
for band insulators} 

We now specialize to band insulators
and evaluate the expectation value of the
non-local operator $\eta$. 
Below,
we will evaluate the expectation value
$\langle \Phi|\eta
|\Phi \rangle
$
for a band insulator. 
Since we can factorize
$
\eta
=
\exp
\left[
\sum_{\mathbf{r}}
\Theta(\mathbf{r})
\rho(\mathbf{r})
\right]
\exp
\left[
-\sum_{\mathbf{r}}
\Theta(\mathbf{r})
\bar{\rho}
\right],
$
we focus on the first factor
(will be denoted by $\eta$ for 
simplicity). 
To this end, we first 
consider
\begin{eqnarray}
&&
\eta =: \xi^{\alpha},
\quad
\xi :=
\exp
\left[
\sum_{\mathbf{r}}
\ln \phi(\mathbf{r})
\rho(\mathbf{r})
\right],
\nonumber \\
&&
\theta(\mathbf{r})
=:
\left[\phi(\mathbf{r})\right]^{\alpha}.
\end{eqnarray}
Note that in real space, 
\begin{eqnarray}
\xi \psi^{\dag}_i(\mathbf{r}) 
= 
\phi(\mathbf{r}) 
\psi^{\dag}_i(\mathbf{r})\xi  
\end{eqnarray}
In the momentum basis, 
\begin{eqnarray}
\xi \psi^{\dag}_i(\mathbf{k})
\!\!&=&\!\!
\sqrt{V}^{-1} \sum_{\mathbf{r}}
\psi^{\dag}_i(\mathbf{r})
\phi(\mathbf{r})
e^{+ {i}\mathbf{r}\cdot \mathbf{k}}
\xi.
\end{eqnarray}
For our choice of $\theta(\mathbf{r})$
and $\phi(\mathbf{r})$,
\begin{eqnarray}
\xi \psi^{\dag}_i(\mathbf{k})
\!\!&=&\!\!
\Big[
+2 d
\psi^{\dag}_{i}(\mathbf{k})
\nonumber \\
&&
\qquad 
-
\sum_{\mu} 
\left(
\psi^{\dag}_{i}(\mathbf{k}+\hat{\mathbf{k}}_{\mu})
+
\psi^{\dag}_{i}(\mathbf{k}-\hat{\mathbf{k}}_{\mu})
\right)
\Big]
\xi
\nonumber \\
\!\!&=:&\!\!
\sum_{\mathbf{k}} 
\psi^{\dag}_i(\mathbf{k}')
T(\mathbf{k}',\mathbf{k}), 
\end{eqnarray}
where
$
\hat{\mathbf{k}}_{\mu}= 
(2\pi) 
\hat{\mathbf{n}}_{\mu}/L
$. 
The matrix $T(\mathbf{k}',\mathbf{k})$ can be viewed 
as a discretized Laplacian (tight-binding Hamiltonian)
on the dual lattice. 

In the basis which diagonalizes
$\mathcal{H}(\mathbf{k})$, 
\begin{eqnarray}
\xi \chi^{\dag}_{\hat a}(\mathbf{k})
\xi^{-1} 
\!\!&=&\!\!
\sum_\mathbf{k} 
\chi^{\dag}_{\hat b}(\mathbf{k}')
T(\mathbf{k}',\mathbf{k})
\langle u^-_{\hat b}(\mathbf{k}') | 
u^{-}_{\hat a}(\mathbf{k}) \rangle
\nonumber \\
\!\!&=:&\!\!
\sum_\mathbf{k} 
\chi^{\dag}_{\hat b}(\mathbf{k}')
\tilde{T}_{\hat{b} \hat{a}}(\mathbf{k}',\mathbf{k}).
\end{eqnarray}
The matrix
$\tilde{T}_{\hat{a} \hat{b}}(\mathbf{k},\mathbf{k}')$
(or 
$\tilde{T}^{\dag}_{\hat{a} \hat{b}}(\mathbf{k},\mathbf{k}')$)
is the properly defined 
``Laplacian'' in 
the ``curved'' momentum space.

The expectation value of the non-local 
operator is then given by 
\begin{eqnarray}
\langle \Phi|\xi
|\Phi \rangle
\!\!&=&\!\!
\mathrm{Det}\,
\left[
\tilde{T}_{\hat{a} \hat{b}}(\mathbf{k},\mathbf{k}')
\right], 
\nonumber \\
\langle \Phi|\eta
|\Phi \rangle
\!\!&=&\!\!
\mathrm{Det}\,
\left[
\tilde{T}^{\alpha}_{\hat{a} \hat{b}}(\mathbf{k},\mathbf{k}')
\right]. 
\label{result}
\end{eqnarray}
where $\mathrm{Det}$
is taken over the hatted indices as well
as the momentum. 
We have thus related
the geometric mean of the position operator squared
to the spectral geometry defined 
for the Laplacian 
in the curved momentum space.

\subsubsection{path-integral representation}

For simplicity, we assume there is only one band occupied.
In this case, 
\begin{eqnarray}
&&
\tilde{T}_{\hat{a} \hat{b}}(\mathbf{k},\mathbf{k}')
:=
T(\mathbf{k},\mathbf{k}')
\langle u^{-}_{\hat a}(\mathbf{k}) | 
u^{-}_{\hat b}(\mathbf{k}') \rangle
\nonumber \\
\!\!&=&\!\!
2d
\delta_{\mathbf{k}', \mathbf{k}} 
-
\sum^d_{\mu=1}
\left[
\delta_{\mathbf{k}', \mathbf{k}+\hat{\mathbf{k}}_{\mu}}
U_{\mu}(\mathbf{k}) 
+
\delta_{\mathbf{k}', \mathbf{k}-\hat{\mathbf{k}}_{\mu}}
U^{\dag}_{\mu}(\mathbf{k}) 
\right] 
\nonumber \\
\!\!&=: &\!\!
2dI - X, 
\end{eqnarray}
where we have assumed two bands $a=\pm 1$,
and we fill the lower band, $\hat{a}=-$.
The link variable
$U_{\mu}(\mathbf{k}) =\langle u (\mathbf{k}+\hat{\mathbf{k}}_{\mu}) | u(\mathbf{k}) \rangle$
is decomposed into
the amplitude (hopping) and phase (gauge field) parts as
\begin{eqnarray}
t_{\mu}(\mathbf{k})
&=&
|\langle u (
\mathbf{k}+\hat{\mathbf{k}}_{\mu}) | u(\mathbf{k}) \rangle|,
\nonumber \\
e^{A_{\mu}(\mathbf{k})}
&=&
\frac{\langle u (
\mathbf{k}+\hat{\mathbf{k}}_{\mu}) | 
u (\mathbf{k}) \rangle}
{|\langle u (
\mathbf{k}+\hat{\mathbf{k}}_{\mu}) | 
u (\mathbf{k}) \rangle|}.
\end{eqnarray}
For an infinitesimal $\hat{\mathbf{k}}_{\mu}$, they are approximated as
\begin{eqnarray}
&&
t_{\mu}(\mathbf{k})
\simeq 
\exp\left(
- \frac{1}{2}g_{\mu\mu}(\mathbf{k}) \Delta k_{\mu} \Delta k_{\mu}
\right),
\nonumber \\
&&
e^{A_{\mu}(\mathbf{k})}
\simeq
\exp\left(
\mathcal{A}_{\mu}(\mathbf{k})\Delta k_{\mu}
\right)
\end{eqnarray}
where $\mu$ is \textit{not} summed,

In order to compute
$\mathrm{Det}\, \tilde{T}
=
\mathrm{Det}\, \left( 2 d I - X\right)$, 
we consider the following \textit{formal} expansion
in terms of $Y=X/(2d)$:
\begin{eqnarray}
 \mathrm{Det}\,(I-Y)
\!\!&=&\!\!
\exp \left[ \mathrm{Tr}\,\ln (I-Y) \right]
\nonumber \\
\!\!&=&\!\!
\exp \left[ -\sum_{n=1}^{\infty}
\frac{1}{n}
\mathrm{Tr}\,
\left(
Y^n
\right)\right].
\end{eqnarray}

For a given $n$, the trace is represented as a sum over paths (loops):
\begin{eqnarray}
\mathrm{Tr}\,
\left(
Y^n
\right)
\!\!&=&\!\!
\left({1\over 2d}\right)^n
\sum_{p}
\sum_{C_p}
\prod_{C_p}^{ |C_p|=n}
U_{\mu}(k) 
\nonumber \\
\!\!&=&\!\!
\left({1\over 2d}\right)^n
\sum_{p}
\sum_{C_p}
\prod_{C_p}^{ |C_p|=n}
t_\mu(\mathbf{k})
e^{A_\mu(\mathbf{k})},
\end{eqnarray}
where $p \in \mathrm{BZ}$ is an initial point for paths,
$C_p$ represents a path starting from $p$ and going back to $p$ 
with $n$ steps,
and $\prod_{C_p}U_{\mu}(k)$ is the path-ordered product.
Thus, the expansion can be interpreted as a
summation over closed loops. 
It should be noted that $\prod_{C_p}U_{\mu}(k)$ 
is actually does not depends on the initial point $p$
at which one starts to draw a path.

Thus,
\begin{eqnarray}
&&
\ln \langle \Phi| \xi |\Phi\rangle
=
\ln \mathrm{Det}\, \tilde{T}
\sim 
\sum_{n=1}^{\infty}{1\over n}
\left({1\over 2d}\right)^n  
\sum_{p}
\sum_{C_p}^{ |C_p|=n}
e^{-S}. 
\nonumber \\
&&
\mbox{where} 
\quad
S =
\frac{1}{2}\int_{C_p} g_{\mu\mu}(k)\Delta k_\mu \Delta k_\mu
-\int_{C_p} \mathcal{A}_{\mu}\Delta k_\mu. 
\end{eqnarray}
We can see that for 
a smaller value of the metric $g_{\mu\nu}$, the non-local operator takes 
a smaller value. 
Especially the trivial metric $g_{\mu\nu}=0$ gives the minimum expectation value of the operator.


\section{Bures metric for topological insulators and superconductors}
\label{case-studies}

In this section,
we explore momentum space geometries and topologies 
of gapped  phases, 
applying general formalism presented in the previous sections.
In particular, we will consider examples of 
topological insulators/superconductors constructed in 
Ref.\ \cite{Ryu09}.
In Ref.\ \cite{Ryu09},
gapped Hamiltonians of Dirac type 
are systematically constructed,
which realize 
topological insulators/superconductors in 
all symmetry classes 
and in all dimensions.
For these, 
we will discuss the momentum space metric
and the second cumulant of the position operators,
by following the Marzari-Vanderbilt formalism, 
and also by following the formalism developed in 
Sec.\ \ref{non-local operator}. 

We will also discuss 
topological invariants made of the Bures metric, 
such as
the Euler class ($d=2n$),
the Pontrjagin class ($d=4n$),
the gravitational Chern-Simons invariant ($d=2n+1$),
and 
the dimensionally continued Euler density ($d=2n+1$).
The dimensionally continued Euler densities have the following forms
\bea
 \Omega^{\mu_1\cdots \mu_{2n} }\wedge E^{\mu_{2n+1}\cdots \mu_{d}}\epsilon_{\mu_1\cdots \mu_{2n}}
\eea
where $ \Omega^{\mu_1\cdots \mu_{2n} }\equiv  \Omega^{\mu_1\mu_2}\wedge \cdots  \Omega^{\mu_{2n-1}\mu_{2n}} $
is the wedge product of the curvature two form and
$ E^{\mu_{2n+1}\cdots \mu_{d}}\equiv E^{\mu_{2n+1}}\wedge \cdots E^{\mu_{d}}$ is the wedge product of the co-frames of the orthonormal frames.
Among the dimensionally continued Euler densities,
there is one which is a topological invariant 
for a given dimension $d$.

Following Ref.\ \cite{Ryu09},
we start with topological Dirac insulators
(Sec.\ \ref{Dirac representatives: generalities})
in $d=2n$- and $d=(2n+1)$- dimensions 
and discuss the general properties
of the momentum space metric.
These Dirac representatives realize 
topological insulators/superconductors 
characterized by an integer topological invariant
for all symmetry classes and for all dimensions
(sometimes called ``primary series'').
We will then focus on specific examples of the 
Dirac representatives in $d=2,3,4$, which include
the QHE, the QSHE, and the 3D $\mathbb{Z}_2$ topological insulator.


\subsection{Dirac representatives: generalities}
\label{Dirac representatives: generalities}

\subsubsection{odd space-time dimensions, $d=2n$}

We start with the $d=2n$-dimensional 
topological Dirac insulator
which is defined in momentum space by 
\begin{eqnarray}
\mathcal{H}^{d=2n}_{(2n+1)}({\bf k})
= 
\sum_{a=1}^{d=2n} k_a \Gamma^a_{(2n+1)}
+m \Gamma^{2n+1}_{(2n+1)}.
\label{eq: class A massive dirac}
\end{eqnarray}
Here,
$k_{a=1,\ldots, d}$
are a component of the $d$-dimensional momentum, 
and 
$\Gamma^{a=1,\ldots, 2n+1}_{(2n+1)}$ 
are $(2^{n}\times 2^{n})$-dimensional hermitian matrices
and satisfy
$
\{\Gamma^a_{(2n+1)},\Gamma^b_{(2n+1)}\}=2\delta_{a,b}.
$
The two eigenvalues of the Dirac Hamiltonian are
\begin{eqnarray}
\pm \lambda(k),
\quad
\lambda(k)
=
\sqrt{k^2 + m^2}, 
\end{eqnarray}
where $k=|\mathbf{k}|$.

The Dirac Hamiltonian (\ref{eq: class A massive dirac})
represents the topological insulator/superconductor in 
class A ($d=2n$),
which is characterized by lack of any discrete symmetry,
and by an integer topological invariant. 
For a given $d=2n$, 
the Hamiltonian (\ref{eq: class A massive dirac})
respects either one of TRS or PHS.
For this reason, 
the Dirac Hamiltonian (\ref{eq: class A massive dirac})
also represents
the topological insulator/superconductor
in class AI ($d=8m$),
class D ($d=8m+2$),
class AII ($d=8m+4$),
class C ($d=8m+6$),
where $m \in \mathbb{Z}$.

The momentum space metric
for 
the Dirac Hamiltonian (\ref{eq: class A massive dirac})
is given,
in the Cartesian
and 
in the radial coordinates,
by
\begin{eqnarray}
ds^2 
\!\!&=&\!\!
\frac{2^n}{8 \lambda^4}
 \left(
 \delta_{ab} \lambda^2 
- k_{a} k_{b} 
\right) dk_a dk_b \cr
\!\!&=&\!\!
\frac{2^n}{8}
\left[
\frac{m^2}{\lambda^4}
dk^2
+
\frac{k^2}{\lambda^2} 
d\Omega_{d-1}
\right],
\label{metric, unregularized}
\end{eqnarray}
respectively. 
Here, 
$ds^2$ is the line element 
in momentum space, 
and 
$d\Omega_{d-1}$
is the line element of the
$(d-1)$-dimensional unit sphere.

Near the origin $k=0$, 
and near $k=\infty$,
the metric is given by
\begin{eqnarray}
ds^2
\simeq 
\left\{
\begin{array}{ll}
\displaystyle
\frac{2^n}{8 m^2} (dk^2+k^2d\Omega_{d-1}),& k\to 0 \\
\\ 
\displaystyle
\frac{2^n}{8} \left(m^2 {dk^2\over k^4}+d\Omega_{d-1}
\right), & k\to \infty.
\end{array}
\right.
\end{eqnarray}
The metric for $k\to 0$ is equivalent to that of a sphere, 
while the metric near $k=\infty$ can be brought into a form,
with $z=1/k$,
$
ds^2
\simeq 
{2^n \over8 }(m^2 dz^2+d\Omega_{d-1})
$.
I.e., the metric of a flat geometry. 
Therefore the momentum space is smoothly capped at $k=0$
while it has a finite radius at $k=\infty$ $(z=0)$.
This shows that the geometry of momentum space,
defined by the Bures metric, is 
a semisphere or a cigar.

\paragraph{regularization}

For the Dirac Hamiltonian
(\ref{eq: class A massive dirac}), 
the behavior of the wavefunctions 
at high energies is non-trivial.
This is signaled by the fact that
the Chern-integer for $d=2n$,
when computed for the Dirac Hamiltonian
(\ref{eq: class A massive dirac}),
is not quantized in integer unit 
but in half-odd integer unit. 
In turn, this is closely related to the fact that
the momentum space geometry detected by the Bures metric is
that of a semisphere.
For these reasons, 
while the Dirac Hamiltonian
(\ref{eq: class A massive dirac})
is capable of describing a transition
between two phases with different topological charge (as we tune $m$), 
it does not represent, in a well-defined fashion, 
a topological or a non-topological insulator
by itself
\cite{Ryu09}.

If we wish,
such non-trivial behavior of the wavefunction can be regularized, 
by modifying the Bloch Hamiltonian
at large momentum,
for example,
by replacing the constant mass by 
\begin{eqnarray}
m \to  \widetilde{m}(k) = m - C k^2
\label{eq mass regularized}
\end{eqnarray}
where $C$ is some constant. 
With the regularization,
the $n$-th Chern integer is quantized in integer unit,
$\mathrm{Ch}_{n}= 1,0$
for $\mathrm{sgn}\, m = \pm \mathrm{sgn} C$,
respectively
($d=2n$). 

The metric for the regularized Dirac Hamiltonian is given by 
\begin{eqnarray}
ds^2 
\!\!&=&\!\!
\frac{2^n}{8}
\left[
\frac{1}{\lambda^2}
dk_a dk_a
+
\frac{ \left(4mC - 1 \right)}{\lambda^4}
k_ak_b
dk_a dk_b
\right]
\nonumber \\
\!\!&=&\!\!
\frac{2^n}{8 }
\left[
\frac{(m + C k^2)^2}{\lambda^4}
 dk^2 
+
\frac{k^2}{\lambda^2} d\Omega_{d-1}
\right],
  \label{metric-eqn-even-reg}
\end{eqnarray}
where $\lambda(k)=\sqrt{k^2 +  \widetilde{m}^2(k)}$. 

Near the origin $k=0$, the metric
shows the same asymptotic behavior,
$ds^2 \simeq \frac{2^n}{8 m^2}(dk^2+k^2d\Omega_{d-1})$, 
as in the unregularized case, 
while near $k=\infty$,
\bea
ds^2
\!\!&\simeq &\!\!
{2^n \over8 C^2}({dk^2\over k^4}+{1\over k^2}d\Omega_{d-1})
\nonumber \\
\!\!&=&\!\!
{2^n \over8 C^2}( dz^2+z^2d\Omega_{d-1}).
\eea
Therefore momentum space is smoothly capped both at 
$k=0$ and at $k=\infty$ $(z=0)$.
This shows that the topology of momentum space is a sphere.

It is also interesting to see the difference between
the regularized and unregularized models at criticality. 
In the limit $m\to 0$, 
the metric for the unregularized model (\ref{metric, unregularized})
reduces to 
$ds^2 =
\frac{2^n}{8}
d\Omega_{d-1}
$, i.e, $g_{kk}=0$. 
On the other hand, for the regularized model,
(\ref{metric-eqn-even-reg}),
the metric reduces
\begin{eqnarray}
ds^2 
\!\!&\to &\!\!
\frac{2^n}{8}
\left[
\frac{C^2  dk^2 }{ 
  (1+C^2 k^2)^2
}
+
\frac{d\Omega_{d-1}}{ 1+C^2 k^2}  
\right]. 
\end{eqnarray}
Notice that the coefficient of $d\Omega_{d-1}$ does
not vanish at $k=0$.
Therefore, 
the topology of momentum space 
changes to create a new ``hole'' at the origin.

Alternatively, one can put the theory on a lattice
to regularize. This amounts to consider the following
lattice version of the Dirac Hamiltonian: 
\begin{eqnarray}
\mathcal{H}^{d=2n}_{(2n+1)}({\bf k})
=
\sum_{a=1}^{d=2n} \sin k_a \Gamma^a_{(2n+1)}
+
f(\mathbf{k})
\Gamma^{2n+1}_{(2n+1)}.
\label{lattice Dirac}
\end{eqnarray}
where
$k_a \in (-\pi,+\pi]$,
and 
${f}({\bf k})
=
(m+d) - 2 C \sum_{a=1}^{d=2n} \cos k_a$.
The eigenvalues are given by  
$\pm \lambda({\bf k})$ with
\begin{eqnarray}
 \lambda({\bf k})
 \!\!&=&\!\!
 \sqrt{ \sum\nolimits_a (\sin k_a)^2 + {f}^2({\bf k})}.
\end{eqnarray}
By expanding (\ref{lattice Dirac}) near $\mathbf{k}\sim 0$, 
one recovers the Dirac Hamiltonian in the continuum,
(\ref{eq: class A massive dirac})
and 
(\ref{eq mass regularized}).
The lattice Dirac Hamiltonian 
(\ref{lattice Dirac})
will be used 
to discuss the expectation value of the 
non-local operator $\eta$. 
The metric in the Cartesian coordinates is given by
\begin{eqnarray}
g_{ab}({\bf k})
\!\!&=&\!\!
\frac{2^n}{8 \lambda^2}
 \left[
 \cos k_a \cos k_b \delta_{ab}
+ 4 C^2 \sin k_a \sin k_b
\right]
\nonumber \\
&&
-
\frac{2^n}{8 \lambda^4}
\sin k_{a}\sin k_b
\nonumber \\
&&
\times
\left[ \cos k_a + 2 C {f}({\bf k}) \right]
\left[ \cos k_{b} + 2 C {f}({\bf k}) \right],
\end{eqnarray}
where repeated indices are not summed over. 
Observe that the volume element $\sqrt{g}$ can be zero,
say, at $k_a=\pm \pi/2$.

\subsubsection{even space-time dimensions, $d=2n-1$}

A Dirac Hamiltonian in even space-time dimensions ($d=2n-1$)
can be obtained from (\ref{eq: class A massive dirac})
by formally replacing one component of the momentum
by a mass term, $k_{2n}\to m_1$ (dimensional reduction),
\begin{eqnarray}
\mathcal{H}^{d=2n-1}_{(2n+1)}({\bf k})
\!\!&=&\!\!
\sum_{a=1}^{d=2n-1} k_a \Gamma^a_{(2n+1)}
\nonumber \\
&&
+m_{1} \Gamma^{2n}_{(2n+1)}
+m_{2} \Gamma^{2n+1}_{(2n+1)}.
\label{eq: class AIII massive dirac}
\end{eqnarray}

When one of the masses is set to zero,
the Hamiltonian anticommutes with $\Gamma^{2n,2n+1}_{(2n+1)}$,
and this is a topological insulator in class AIII,
i.e., there is chiral (sublattice) symmetry
In addition to this chiral symmetry, 
for a given $d=2n+1$, 
the Hamiltonian (\ref{eq: class AIII massive dirac})
respects either one of TRS or PHS.
For this reason, 
the Dirac Hamiltonian (\ref{eq: class AIII massive dirac})
also represents
the topological insulator/superconductor
in class BDI ($d=8m+1$),
class DIII ($d=8m+3$),
class CII ($d=8m+5$),
class CI ($d=8m+7$),
where $m\in \mathbb{Z}$.

With the Dirac representatives
(\ref{eq: class A massive dirac})
and
(\ref{eq: class AIII massive dirac}),
we cover all topological insulators/superconductors
labeled by an integer topological invariant
(called ``primary series'').
Topological insulators/superconductors
characterized by an binary topological invariant ($\mathbb{Z}_2$)
or an even integral topological invariant ($2\mathbb{Z}$)
can be derived from these by dimensional reduction. 
We will not give an exhaustive study of the momentum space metric
for these ``descendents'' (i.e.,  topological insulators/superconductors
characterized either by $\mathbb{Z}_2$ or $2\mathbb{Z}$ topological invariant).
Rather, we will focus on the QSHE (class AII in $d=2$)
and the 3D $\mathbb{Z}_2$ topological insulator (class AII in $d=3$)
(see below).

The two eigenvalues of the Dirac Hamiltonian (\ref{eq: class AIII massive dirac}) is 
\begin{eqnarray}
\pm \lambda(k),
\quad
\lambda(k)
=
\sqrt{k^2 + m^2_{1}+m^2_{2}}.
\end{eqnarray}
The calculations of the metric, etc., go parallel with 
the case of $d=2n$. 
The quantum metric is given by 
\begin{eqnarray}
g_{ab}({\bf k})
\!\!&=&\!\!
\frac{2^n}{8 \lambda^4}
 \left(
 \delta_{ab} \lambda^2 
- k_{a} k_{b} 
\right). 
\end{eqnarray}
This is the same metric as that of the $d=2n$ case.
Therefore the geometry of momentum space
is a cigar.

As before,
non-trivial behaviors of the Bloch wavefunction
can be regularized, 
by replacing the constant mass by 
\begin{eqnarray}
m_{1,2} 
\!\!&\to &\!\!
\widetilde{m}_{1,2}(k) = m_{1,2} - C_{1,2} k^2. 
\end{eqnarray}
With the regularization, the metric can be written 
\begin{eqnarray}
ds^2
\!\!&=&\!\!
\frac{ 2^n}{8\lambda^2}
dk_a dk_a 
+
\frac{ 2^n}{8\lambda^2}
k_a k_b 
\nonumber \\
&&
\times 
\left[
4 |\vec{C}|^2 |\vec{m}|^2
-
\big(
2 \vec{m}\cdot \vec{C} -1 
\big)^2
\right]
dk_a dk_b
\nonumber \\
\!\!&=&\!\!
\frac{2^n}{8 \lambda^4}
\big[
\lambda^2 
+
4 |\vec{C}|^2 |\vec{m}|^2 k^2
-
k^2 
\big(
2 \vec{m}\cdot \vec{C} -1 
\big)^2
\big]
dk^2
\nonumber \\
&&
+
\frac{2^n}{8 \lambda^2}
 k^2 d\Omega_{d-1}
\end{eqnarray}
where 
$\lambda(k)=\sqrt{k^2 +  \widetilde{m}\cdot  \widetilde{m}}$,
and we have introduced a vectorial notation, for instance, 
$ \vec{C}\cdot \vec{m}=C_1 m_1 +C_2 m_2$, 
$|\vec{C}|^2 = \vec{C}\cdot \vec{C}$,
etc.

The structures of the metric at $k=0$ and $k=\infty$ are the same as those of $d=2n$ case.
Therefore, the topology of the momentum space is a sphere.

\subsection{example in $d=2$: the IQHE (class A)}

Consider three mutually anticommuting,
hermitian matrices
$\Gamma^{a=1,2,3}_{(3)}=
\{
\sigma_{x},
\,
\sigma_{y},
\,
\sigma_{z}
\},
$
where $\sigma_{x,y,z}$ are
the $2\times 2$ Pauli matrices. 
We consider the following 
(unregularized) $d=2$-dimensional 
massive Dirac Hamiltonian, 
\begin{eqnarray}
\mathcal{H}^{d=2}_{(3)}({\bf k})
\!\!&=&\!\!
k_x \sigma_x 
+
k_y \sigma_y
+
m \sigma_z.
\label{QHE Dirac Hamiltonian}
\end{eqnarray}
This is a class A Hamiltonian.

Assuming $m>0$,
the Berry connection 
and the Berry curvature 
are given by 
\begin{eqnarray}
&&
A_x ({\bf k}) =
+\frac{{i} k_y }{2 \lambda \left(\lambda +m\right)},
\quad
A_y ({\bf k}) =
-\frac{ {i} k_x }{2 \lambda \left(\lambda +m\right)},
\nonumber \\
&&
F_{xy}\left({\bf k} \right) =
\partial_{k_x} A_y - \partial_{k_y} A_x
= -\frac{{i} m }{ 2 \lambda^3 }.
\end{eqnarray}
The Chern number 
is non-zero, 
\begin{eqnarray}
\mathrm{Ch}_{1}
=
\frac{{i}}{2 \pi } \int d^2k\, F_{xy}
=
\frac{1}{2}
\frac{m}{|m|}, 
\end{eqnarray}
which is nothing but 
the Hall conductivity $\sigma_{xy}$.
This model can be realized in the low energy limit of 
the honeycomb lattice model 
introduced by Haldane \cite{haldane-honey}. 
While the Chern integer in the unregularized model
(\ref{QHE Dirac Hamiltonian}) in quantized in half-integer unit,
with regularization, 
it is quantized in integer unit.

\paragraph{the Marzari-Vanderbilt cumulant}

We now discuss
quantities constructed from
the momentum space metric.
Let us first discuss 
the Marzari-Vanderbilt cumulant $\Omega_{I,\mu\nu}$.
Plugging the concrete form of the metric, we obtain,
for the diagonal component, 
\begin{eqnarray}
&& \Omega_{I,xx}
=
\Omega_{I,yy} 
\\
&&
=
{1\over 16\pi}
\Bigg[
{(2Cm-1)k^2-2m^2\over 2(C^2k^4-(2Cm-1)k^2+m^2)}
\nonumber \\
&&
+
{2Cm+1 \over\sqrt{4Cm-1}}\arctan\left({2C^2k^2-2Cm+1\over\sqrt{4Cm-1}}\right)
\Bigg]\Bigg|_{k=0}^{k=\infty}\cr
&&\to
{1\over16\pi}
 \cr
&&
+{1\over 32}
{2Cm+1\over \sqrt{4Cm-1}}
\Bigg[
1-\arctan\left({-2Cm+1\over\sqrt{4Cm-1}}\right)
\Bigg] 
\nonumber \cr
\end{eqnarray}
while the off-diagonal component is vanishing $\Omega_{I,xy}=0$.

\begin{figure}[tb]
\begin{center}
    \includegraphics[height=.25\textwidth]{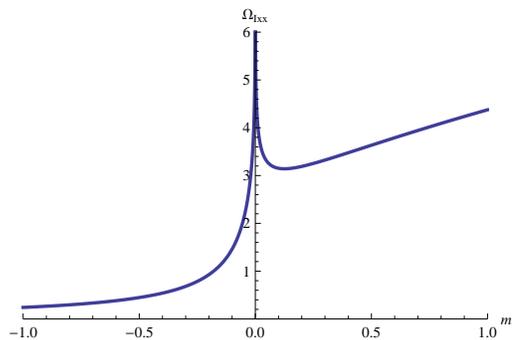}
  \caption{
The Marzari-Vanderbilt cumulant $\Omega_{I,xx}$ 
for the 2D Dirac model (\ref{QHE Dirac Hamiltonian})
as a function of the mass parameter $m$.
The mass term is regularized by replacing $m$
in (\ref{QHE Dirac Hamiltonian})
by $m\to \tilde{m}(k)= m- C k^2$,
and we have set $C=1$.
\label{2dz-cum-reg-fig}}
\end{center}
\end{figure}

The behavior $\Omega_{I,xx}$ as a function of the mass parameter $m$ 
is shown in Fig.\ \ref{2dz-cum-reg-fig}. 
Observe that the plot is {\it not} symmetric 
about $m=0$: the cumulant for the positive mass $m>0$
is larger than for the negative mass $m<0$.
This shows that the electrons tend to be delocalized in the topological phase
($m> 0$ in this case).
In particular, 
the cumulant goes to zero
 as $m\to-\infty$, while it increases for positive $m$.
%
This behavior should be compared with 
the lattice calculation 
in Refs.\ \cite{ThonhauserVanderbilt06,MaChenFanLiu}, 
where the cumulant remains finite in
the topological phase. 
Had we not regularized the Dirac model (\ref{QHE Dirac Hamiltonian}),
$\Omega_{I,xx}$ as a function of the mass parameter $m$ 
would be symmetric with respect to $m=0$. 
At the critical point $m=0$, the cumulant diverges logarithmically 
\begin{eqnarray}
\Omega_{I,xx}\sim \log m.
\end{eqnarray}



\paragraph{topological invariant}

The topological number for a closed 2 dimensional manifold is the Euler number
\begin{eqnarray}
\chi
=
\frac{1}{4\pi} \int d^2k  \sqrt{g} R, 
\end{eqnarray}
which is equal to $2-2g-b$ for a smooth manifold with genus $g$ and boundary number $b$.

As we mentioned above, the momentum space is a cigar and non-compact if we do not introduce a regulator.
With the regularization $m \to \widetilde{m}(k) = m - C k^2$, the topology of the momentum space is
 a sphere and so the Euler number is $\chi=2$.
Note that when $\mathrm{sgn}(m)=-\mathrm{sgn}(C)$,
the metric  (\ref{metric-eqn-even-reg})  has a coordinate singularity at $k^2=-m/C$.

\paragraph{second cumulant: the geometric mean}

We now study the expectation value of the non-local operator
$\eta$ for a topological Dirac insulator in class A. 
Here, we considered the lattice regularized model
(\ref{lattice Dirac})
with $d=2$.
The Dirac model describes
the chiral $p$-wave superconductor ($p$-wave SC)
on the square lattice defined by the tight-binding Hamiltonian 
\begin{eqnarray}
&&
H
=
\sum_{\mathbf{r}}
\psi(\mathbf{r})^{\dag}
\left(
\begin{array}{cc}
t & \Delta\\
-\Delta  & -t
\end{array}
\right)
\psi(\mathbf{r}+\hat{\mathbf{x}})
+\mathrm{h.c.}
\nonumber \\
&&
\qquad
+
\psi(\mathbf{r})^{\dag}
\left(
\begin{array}{cc}
t & {i}\Delta\\
{i}\Delta  & -t
\end{array}
\right)
\psi(\mathbf{r}+\hat{\mathbf{y}})
+\mathrm{h.c.}
\nonumber \\
&&
\qquad
+
\psi(\mathbf{r})^{\dag}
\left(
\begin{array}{cc}
\mu & 0\\
0  & -\mu
\end{array}
\right)
\psi(\mathbf{r}),
\label{lattice p-wave}
\end{eqnarray}
where 
the multi-component fermion annihilation 
operator at site $\mathbf{r}$,
$\psi(\mathbf{r})$, 
is given in terms of 
the electron annihilation operators
with spin up and down,
$c_{\uparrow,\downarrow}^{\ }(\mathbf{r})$,
as
$\psi^{T}(\mathbf{r})=
(c_{\uparrow}^{\ }(\mathbf{r}), c_{\downarrow}^{\dag}(\mathbf{r}))$,
and we take
$t=\Delta=1$ and $\mu=m+2$.
The chiral 
$p$-wave SC
has been discussed in the context of 
superconductivity in a ruthenate 
and paired states in the fractional quantum Hall effect
\cite{Read00, volovik, Senthil98, goryo, morita, Hatsugai02}. 
There are four phases separated by three quantum critical points
at $\mu=0,\pm 4$, which are labeled by
the Chern number as
$\mathrm{Ch}=0$ $( |\mu| > 4)$,
$\mathrm{Ch}=-1$ $(-4 < \mu < 0)$, and
$\mathrm{Ch}=+1$ $( 0 < \mu < +4)$.
The non-zero Chern number implies the IQHE in the spin transport \cite{Senthil98}.


\begin{figure}[tb]
\begin{center}
    \includegraphics[height=.3\textwidth]{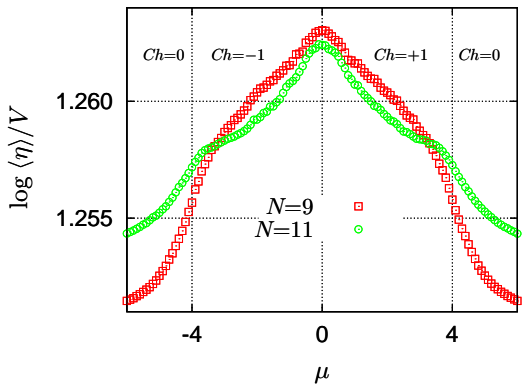}
  \caption{
The expectation value
of the non-local operator
$V^{-1}
\ln\, \langle \Phi|\eta
|\Phi \rangle$
for the lattice Dirac Hamiltonian
(lattice chiral $p$-wave superconductor) (\ref{lattice p-wave})
as a function of $\mu$. 
\label{dual_laplace}
}
\end{center}
\end{figure}

In Fig.\ \ref{dual_laplace},
we plot the expectation value
of the non-local operator, 
$\alpha^{-1}
\ln\, \langle \Phi|\eta
|\Phi \rangle$,
obtained by numerically diagonalizing 
the Laplacian operator in momentum space 
$\tilde{T}_{\hat{a} \hat{b}}(\mathbf{k},\mathbf{k}')$,
$
V^{-1}
\ln\, \langle \Phi|\eta
|\Phi \rangle
=
V^{-1}
\ln\,  
\mathrm{Det}\,
\big[
\tilde{T}_{\hat{a} \hat{b}}(\mathbf{k},\mathbf{k}')
\big]
$. 
The behavior of 
$V^{-1} \ln \langle \Phi|\eta
|\Phi \rangle$
as a function of $\mu$ is 
qualitatively similar to the Marzari-Vanderbilt cumulant:
(i) it is larger in the topological phase
($-4 < \mu < 4$)
than in the non-topological phases
($|\mu| > 4$), 
and
(ii)
near the critical points $\mu=\pm 4, 0$, 
it shows a peak structure. 
Due to the system size limit 
accessible by numerics,
it is admittedly difficult to see
if 
$V^{-1} \ln \langle \Phi|\eta
|\Phi \rangle$
develops a singularity 
at the critical points. 

Some technical comments are in order. 
Computing 
$V^{-1} \ln \langle \Phi|\eta
|\Phi \rangle$
by numerically diagonalization the momentum space Laplacian,
and then summing over logarithm of eigenvalues
can be quite tricky, 
since some eigenvalues are very close to zero.
In order to obtain a meaningful result, 
we can modify (or regularize) 
$\phi(\mathbf{r})$ as
\begin{eqnarray}
\phi(\mathbf{r}) 
\!\!&=&\!\!
-2 
\sum^d_{\mu=1} 
\cos 
\frac{2\pi n_{\mu}}{N} 
+2d 
\nonumber \\
\!\!&\to &\!\!
-2 
\sum^d_{\mu=1} 
\cos \frac{2\pi (n_{\mu}+\delta n_{\mu})}{N} 
+2(d + \delta d).
\end{eqnarray}
Here, adding $\delta d\ll d$ shifts 
all eigenvalues $\varepsilon_i$ of the Laplacian,
$\varepsilon_i\to \varepsilon +2\delta d$,
thereby prevents having an eigenvalue which is too close to zero.
The expectation value
$\ln\,  
\mathrm{Det}\,
\big[
\tilde{T}_{\hat{a} \hat{b}}(\mathbf{k},\mathbf{k}')
\big]$
can be evaluated for small but finite $\delta d$,
and the results can be interpolated to $\delta d\to 0$. 
Adding $0< \delta n_{\mu}< 1$ amounts
shifting the ``location'' of the non-local operator
at which it is inserted;
the function $\phi(\mathbf{r})$ is centered at 
a site $\mathbf{r}=0$,
but alternatively, one can define 
$\phi(\mathbf{r})$ in such a way that it is centered
at the center of a plaquette (i.e., a dual site),
$\delta n_{\mu}=1/2$.
Or, more generally, one can take the center of
$\phi(\mathbf{r})$ anywhere in
a plaquette,
$0<\delta n_{\mu} <1$.
In Fig.\ \ref{dual_laplace},
the expectation value 
$\ln\,  
\mathrm{Det}\,
\big[
\tilde{T}_{\hat{a} \hat{b}}(\mathbf{k},\mathbf{k}')
\big]$
is averaged over different centers (different choice of $0<\delta n_{\mu} <1$)
to obtain smooth curves.

\subsection{example in $d=2$: the QSHE (class AII)}

Consider five mutually anticommuting,
hermitian matrices
\begin{eqnarray}
\Gamma^{a=1,\ldots,5}_{(5)}=
\{
\alpha_{x},
\,
\alpha_{y},
\,
\alpha_{z},
\,
\beta,
\, 
-{i} \beta \gamma^5
\},
\label{eq: 4x4 gamma matrices}
\end{eqnarray}
where we are using the Dirac representation, 
\begin{eqnarray}
\alpha_i
=\left(\begin{array}{cc}
0 & \sigma_i\\
\sigma_i & 0
\end{array}\right),
~
\beta=\left(\begin{array}{cc}
1 & 0\\
0 & -1\end{array}\right),
~
\gamma^{5}=\left(\begin{array}{cc}
0 & 1\\
1 & 0\end{array}
\right).
\end{eqnarray}
Observe that
$
\alpha_x
\alpha_y
\alpha_z
\beta 
= -{i} \beta\gamma_5
$.
We will use these gamma matrices to construct Dirac
representative for $d=2,3,4$ below.

Let us consider the following $d=2$-dimensional
Dirac Hamiltonian.
\begin{eqnarray}
\mathcal{H}^{d=2}_{(5)}({\bf k})
\!\!&=&\!\!
k_x \alpha_1
+
k_y \alpha_2
+
\alpha \gamma_5
+
\widetilde{m}(k) (-{i}\beta\gamma_5),
\nonumber \\
\widetilde{m}(k)
\!\!&=&\!\!
m - Ck^2. 
\label{z2-2d hamiltonian}
\end{eqnarray}
This Hamiltonian is time reversal invariant,
\bea
T\mathcal{H}^{*}(-\mathbf{k})T^{-1}=\mathcal{H}(\mathbf{k}),
\quad
T=i\tau_{x}\otimes\sigma_{y}. 
\eea
The Dirac Hamiltonian of type (\ref{z2-2d hamiltonian})
was used in Refs.\ \cite{KaneMele, bernevig06}
to discuss the QSHE.
In the Kane-Mele model on the honeycomb lattice, 
the mass $m$ is generated by a time-reversal
spin-dependent second neighbor hopping, 
whereas the $\alpha$-term arises from the Rashba spin-orbit coupling.
Time reversal invariance is crucial in the QSHE.
When the edge states consist of 
even number of Kramers' pairs of counter propagating 
modes (helical modes),
they can be gapped under perturbations. 
On the other hand, when the number of helical modes is odd,
they cannot be gapped unless time reversal symmetry is broken.

The eigenvalues of the hamiltonian (\ref{z2-2d hamiltonian}) are
\bea
E(k)&=&\pm \lambda^{+}_{k},~\pm \lambda^{-}_{k} \cr
\lambda^{\pm}_{k} &=&\sqrt{(k\pm \alpha)^2+(m - Ck^2)^2}
\eea
Notice that for $\alpha=0$, there is a critical point at $m=0$ and $k=0$.
For $\alpha>0,~m/C>0$, there is a critical point at $m/C=\alpha^2$ and $k=\sqrt{m/C}=\alpha$.
As observed in Refs.\ \cite{OFRM2007,SMOF}, 
there is a metallic phase near $m=0$ for $\alpha\neq0$ sandwiched 
by insulating phases, while
for $\alpha=0$, the metallic phase shrinks to a critical point.


The Bures metric is given by
\bea
&&
ds^2 
\nonumber \\
%
\!\!&=&\!\!
{1\over4}
\left[
\frac{(m+Ck(k-2\alpha))^2}{A_{-}(k)^2}
+
\frac{(m+Ck(k+2\alpha))^2}{A_{+}(k)^2}
\right]
dk^2 \cr
&&+
{1\over4}
\left[
1+
\frac{
k^2-\alpha^2-(m-Ck^2)^2}
{\sqrt{A_+(k) A_{-}(k)}
}
\right]
d\varphi^2.
\label{z2-2d-metric}
\eea
where
\begin{eqnarray}
A_{\pm}(k)
:=
(m-Ck^2)^2+(k\pm \alpha)^2. 
\end{eqnarray}
Observe that
$g_{kk}$ is a non-negative function as seen from (\ref{z2-2d-metric}),
while the behavior of $g_{\varphi\varphi}$ is different for
topologically trivial and topologically non-trivial phases;
$g_{\varphi\varphi}$ is zero at $k= \sqrt{m/C}$ if
$
-{m\over C}+\alpha^2 >0.
$

Let us first discuss several asymptotic behaviors 
of the metric.
We first consider the case of $m\neq0$.
Near the origin $k=0$, the metric is given by 
\bea
ds^2={m^2\over 2(m^2+\alpha^2)^2}(dk^2+k^2d\varphi^2)+\cdots.
\label{QSHE,k=0}
\eea
On the other hand, asymptotic value of $g_{kk}$ $(k\to \infty)$ is ${1\over 2C^2 k^4}$.
By making a coordinate transformation
$z=1/(\sqrt{2}Ck)$,
the metric is given by 
\bea
ds^2=dz^2+z^2d\varphi^2+\cdots.
\label{QSHE,k=inf}
\eea
Both (\ref{QSHE,k=0}) and (\ref{QSHE,k=inf})
are the metric of a sphere, and therefore
we see that the momentum space manifold is smoothly capped both at
the origin and the infinity.

On the other hand, 
when $m=0$ and $\alpha\neq0$,
the metric near the origin $k=0$ is
\bea
ds^2
\!\!&=&\!\!
{C^2\over 2\alpha^2}\left(d(k^2)^2+{(k^2)^3\over\alpha^2}d\varphi^2\right) +\cdots.
\eea
This metric has a singularity at the origin.
This should be compared with the case of $m=0$ and $\alpha=0$,
for which  the metric near the origin is
\bea
ds^2={C^2\over2}dk^2+{1\over2}d\varphi^2.
\eea
The $\varphi$ circle does not shrink to zero at the origin and the metric is analytically continued to
the positive energy region.
Note that the energy becomes zero at $k=0$ and the gap is closed.

\paragraph{the Marzari-Vanderbilt cumulant}

\begin{figure}[tb]
\begin{center}
    \includegraphics[height=.25\textwidth]{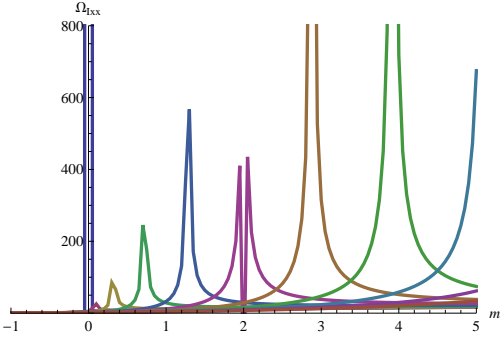}
          \caption{
The Marzari-Vanderbilt cumulant $\Omega_{I,xx} (=\Omega_{I,yy})$ 
for the two-dimensional $\mathbb{Z}_2$ topological Dirac insulator
(\ref{z2-2d hamiltonian})
as a function of the mass parameter $m$.
Here, $C=2$,
and 
$\alpha =0, 0.2, 0.4, 0.6, 0.8, 1, 1.2, 1.4, 1.6$.
from the left to the right.
\label{2dz2cum-New-multi-2}
}
\end{center}
\end{figure}

The Marzari-Vanderbilt second cumulant is calculated numerically in
Fig. \ref{2dz2cum-New-multi-2}.
At the critical point $\alpha=m=0$, the cumulant diverges. 
For finite $\alpha$, the cumulant is finite 
with a peak at at $m=C\alpha^2$,
whose peak value increases as $\alpha$ increases.
In another critical case (pink line in Fig.\ \ref{2dz2cum-New-multi-2}) 
where the radial direction of the momentum space 
shrinks to zero at certain value of $k$, 
the cumulant gets two peaks and at the critical point  
$m=C\alpha^2(=2~\text{in Fig.\ \ref{2dz2cum-New-multi-2}})$, the
cumulant takes small but finite value.
The cumulant changes very rapidly around the critical point $m=C\alpha^2$.
Above this value $\alpha=1$, the cumulant monotonically increases.  
Another important property of the cumulant is that for fixed $\alpha$, the cumulant takes larger value in $m-C\alpha^2>0$.
As studied in \cite{SRFL, SMOF}, $m-C\alpha^2>0$ region is the $\mathbb{Z}_2$ topological superconductor phase.
Therefore, this shows that the electron tends to be delocalized in the topologically non-trivial phase.

\subsection{examples in $d=3$: 
chiral topological insulator (class AIII)
and
$\mathbb{Z}_2$ topological insulator (class AII)}

As in the general discussion before,
we can consider two mass terms in $d=3$, 
\begin{eqnarray}
\mathcal{H}^{d=3}_{(5)}({\bf k})
\!\!&=&\!\!
\sum_{a=1}^3
k_a \Gamma^a_{(5)}
+
m_1 \Gamma^4_{(5)}
+
m_2 \Gamma^5_{(5)}.
\label{eq: dirac 3d}
\end{eqnarray}
When either one of the masses are switched off,
this Hamiltonian belongs to the symmetry class
AIII, or AII. 

As discussed in 
Ref.\ \cite{Qi2008} for class AII and
in Ref.\ \cite{Pavan09} for class AIII, 
non-trivial topology of the Dirac Hamiltonian 
is signaled by the existence of the
$\theta$-term in the effective action for
the linear electromagnetic response
with $\theta=\pi$.
The non-zero (and quantized) $\theta$-term, in turn, 
is related to the topological invariant made of
the Berry connection, 
which is given for
the Dirac model (\ref{eq: dirac 3d}), 
\begin{eqnarray}
\mathrm{CS}_3
\!\!&=&\!\!
\frac{-1}{8\pi^2}
\int d^3 k\, 
\epsilon^{\mu\nu\rho}
\mathrm{tr}\,\Big(
A_{\mu} \partial_{\nu} A_{\rho}
+
\frac{2}{3}
A_{\mu} A_{\nu} A_{\rho}
\Big)
\nonumber \\
\!\!&=&\!\! 
\frac{1}{\pi}
\frac{m_2}{|m_2|}
\mathrm{arctan}
\left[
\frac{
(m^2_2 + m^2_1)^{1/2} - m_1}
{
(m^2_2 + m^2_1)^{1/2} + m_1
}
\right]^{1/2}. 
\end{eqnarray}
In particular, when $m_1=0$,
$\mathrm{CS}_3
=
(1/4) m_{2}/|m_{2}|.
$
As the CS term is determined only modulo 1, 
the fractional part
$(1/4) \mathrm{sgn}(m_2)$ is an intrinsic property.
On a lattice, as there are two-copies of the $4\times4$ Dirac Hamiltonians,
the total Chern-Simons invariant is given by 
$\mathrm{CS}_3=({1}/{2})\mathrm{sgn}(m_{2})$,
and $\exp (2\pi{i}\mathrm{CS}_3) =-1$.

Let us now set $m_1=0$, and, as before, 
regularize the mass, 
\begin{eqnarray}
\mathcal{H}^{d=3}_{(5)}({\bf k})
\!\!&=&\!\!
k_x \alpha_1
+
k_y \alpha_2
+
k_z \alpha_3
+
\widetilde{m}(k) (-{i}\beta\gamma_5),
\nonumber \\
\widetilde{m}(k)
\!\!&=&\!\!
m - C k^2.
\label{eq: 3D Dirac}
\end{eqnarray}
This Hamiltonian can be obtained from
the Dirac representative of the QSHE 
by replacing 
$\alpha$ in the two dimensional hamiltonian (\ref{z2-2d hamiltonian})
by $k_z$.

The metric 
for the Dirac model (\ref{eq: 3D Dirac})
is 
\bea
ds^2
\!\!&=&\!\!
\frac{1}{2 [(Ck^2-m)^2+k^2]}
\Big[
{(Ck^2+m)^2\over (Ck^2-m)^2+k^2} dk^2
\nonumber \\
&&
+k^2 (d\theta^2+\sin^2\theta d\phi^2)
\Big], 
\label{eq: metric 3D Dirac} 
\eea
where we used the spherical coordinate, 
$(k_x, k_y, k_z)=(k \sin\theta \cos\phi, k \sin\theta \sin\phi, k\cos\theta)$.

The behavior of the metric at $k=\infty$ is similar to that of the $d=2$ case.
Here, we check the regularity of the metric at the origin. 
For $m\neq0$, the metric is
\bea
ds^2\cong {1\over 2m^2}
\left[ dk^2+k^2(d\theta^2+\sin^2\theta d\phi^2)\right]
\eea
while for $m=0, C\neq 0 $, it is
\bea
ds^2\cong {C^2\over 2}dk^2+{1\over2}(d\theta^2+\sin^2\theta d\phi^2).
\eea
The momentum space is regular for finite $m$, 
while at the critical point $m=0$ there is a hole at the origin $k=0$.
Note that when $\text{sign}(C)=-\text{sign}(m)$, 
there is a coordinate singularity $g_{kk}=0$ at
$Ck^2+m=0$.
This suggests that the cumulant for $\text{sign}(C)=-\text{sign}(m)$ phase would take smaller value than that for
 $\text{sign}(C)=\text{sign}(m)$ phase. We will show this explicitly in the next paragraph. 

\paragraph{the Marzari-Vanderbilt cumulant}

\begin{figure}[tb]
\begin{center}
    \includegraphics[height=.25\textwidth]{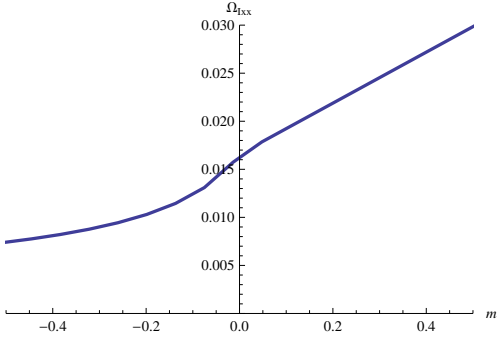}
          \caption{
The Marzari-Vanderbilt cumulant $\Omega_{I,xx} (=\Omega_{I,yy})$ 
as a function of the mass parameter $m$ for 
for the 3D Dirac model (\ref{eq: 3D Dirac})
with $C=2$.
\label{3dz2cum-fig}
}
\end{center}
\end{figure}

The diagonal component of the Marzari-Vanderbilt cumulant
$\Omega_{I,xx} (= \Omega_{I,yy})$
for the 3D Dirac Hamiltonian (\ref{eq: 3D Dirac}),
obtained by numerically integrating the metric 
over three dimensional momentum space, 
is shown in Fig.\ \ref{3dz2cum-fig}.
The off-diagonal component is identically zero,
$\Omega_{I,xy} = 0$.
The cumulant does not diverge at $m=0$,
while  its first derivative with respect to the mass $m$
changes discontinuously: 
\bea
\Omega_{I,xx}
=
\begin{cases}
\displaystyle
{5\over 48\pi C}+
{5\over 24\pi}m+\cdots,  & m\to -0 \cr
\cr 
\displaystyle
{5\over 48\pi C}+
{1\over 12\pi}m+\cdots.   & m\to+0 \cr
\end{cases}
\eea
As mentioned above, the negative mass $(m<0)$ phase has a coordinate singularity $(g_{kk}=0)$
and the cumulant becomes smaller compared to the positive mass phase.
The cumulant gets larger 
for large positive $m$. 
However, since this would come from the UV region of the integration,
it would not have physical significance.


\paragraph{topological invariants}

In $d=3$ momentum space,
one can think of two invariants.
Both of them are related to a Chern-Simons characteristics class,
derived from the Pontrjagin, and
the Euler characteristics classes,
respectively. 

The Chern-Simons invariant derived from the
Pontrjagin class is the gravitational Chern-Simons term,
which was discussed, for example, 
in Refs.\ \cite{Deser, Read00}, 
and is given by 
\begin{eqnarray}
I_{\mathrm{GCS}}
\!\!&=&\!\!
\frac{-1}{8\pi^2} 
\int d^3k\, 
\mathrm{tr}\, 
\Big(
 \omega  d\omega
+
\frac{2}{3} \omega^3 
\Big), 
\end{eqnarray}
where $\omega$ is a connection one form. 
This invariant turns out to be zero,
since the integrand vanishes identically
for 
the metric (\ref{eq: metric 3D Dirac}).

Another invariant is 
the dimensionally continued Euler number.
In three-dimensions,  
dimensionally continued Euler number can be written as 
\begin{eqnarray}
\chi
\!\!&=&\!\!
\frac{1}{2(4\pi)^2}
\int 
\epsilon_{abc}
\Omega^{ab}
\wedge e^{c} 
\end{eqnarray}
where
$
\epsilon_{abc}
\Omega^{ab}
\wedge e^{c} 
=
\epsilon_{abc}
R^{ab}{ }_{kl}
e^{k}
\wedge
e^{l}
\wedge e^{c} 
=
2R^{ab}{ }_{ab}
\sqrt{g} d^3 k
$.
For the unregularized model,
the Dirac (\ref{eq: 3D Dirac}) with $C=0$, 
the dimensionally continued Euler number is given by 
\begin{eqnarray}
\chi
= 
\frac{3\sqrt{2}}{4^2}. 
\end{eqnarray}
It is straightforward to check that this result still holds for the regularized model where $m_2$ is replaced by
$\widetilde{m}_2(k)=
m_2 - C_2k^2. $
When $m_1=0$ and $m_2=1$
with $C_2=1$, the integral can be evaluated 
analytically, and gives the same result as above.

\subsection{example in $d=4$}

The gamma matrices (\ref{eq: 4x4 gamma matrices})
can be used to construct
a $d=4$ dimensional Dirac Hamiltonian, 
\begin{eqnarray}
\mathcal{H}^{d=4}_{(5)}({\bf k})
=
\sum_{a=1}^4
k_a \Gamma^a_{(5)}
+
m \Gamma^5_{(5)}.
\label{eq: dirac 4d}
\end{eqnarray}
The Hamiltonian (\ref{eq: dirac 4d})
is a topological insulator in class A
with the non-zero second Chern number 
constructed with the Berry curvature 
\cite{Golterman, Qi2008}.
The Hamiltonian (\ref{eq: dirac 4d})
respects time-reversal symmetry which squares to be -1,
and hence belongs to symmetry class AII. 
The 3D Dirac Hamiltonian (\ref{eq: dirac 3d}),
which is a $\mathbb{Z}_2$ topological insulator in class AII,
can be obtained from (\ref{eq: dirac 4d}) by dimensional reduction. 
\cite{Qi2008}.

\paragraph{topological invariants}

The Euler number calculated by using the Bures metric is
\begin{eqnarray}
\chi
\!\!&=&\!\!
\frac{1}{128\pi^2}
\int d^4k\,
\sqrt{g} 
\epsilon^{klmn}
\epsilon^{abcd}
R_{klab}
R_{mncd}
=1. ~~~
\end{eqnarray}
On the other hand, 
the Pontrjagin number
(the gravitational instanton term, or the signature $\tau$) 
is identically zero: 
\begin{eqnarray}
P
\!\!&=&\!\!
\frac{1}{32\pi^2 } 
\int d^4k\,
\sqrt{g} 
R_{mnlr}
\epsilon^{mnab}
R^{lr}{ }_{ab}
=0. 
\end{eqnarray}

\section{Concluding remark 
and discussion on interaction effects}

In this paper, we studied several properties of 
quantum ground states in momentum space by using the Bures metric.
The Berry gauge potential and the Bures metric naturally appear in 
the expectation value of the non-local operator which measures the charge distributions.
Namely, the Berry gauge potential measures 
polarization in real space and the Bures metric measures the second cumulant.
These are the first two expansions of the generating function (\ref{gene-fucn}).

We introduced the distance square operator in real space which corresponds to 
the Laplacian in the curved momentum space.
In the path integral description, this operator is given by the summation of geodesic lengths and phases of all closed loops 
in momentum space.

These operators, the second cumulant and the distance square operators, are applied to
analyze some topological insulators.
In certain phases, the Bures metrics possess the coordinate singularities where some components of the metric vanish.
As a result, the cumulants take smaller values in those phases compared to other topologically non-trivial phases.
These operators also capture the singular behaviors of the quantum phase transitions.

We also calculated several topological quantities such as the Euler number by using the Bures metric.
These are analogous of the Chern number for the Berry gauge potential, which does characterize topological phases
and has physical meaning. At this point, it is not very clear if the topological numbers for the Bures metric directly 
related to some observational quantities. The Euler number, for instance, measures the topologies of the momentum spaces.
Naively these are spheres or tori depending on regularizations. However, we have seen in the case of the $2d$ $\mathbb{Z}_2$ topological insulator that the topology could be non-trivial: one cycle shrinks to zero at finite momentum $k$.  Also, we have seen that there are several types of singularities in the momentum space geometries
for both of the $\mathbb{Z}$ and  $\mathbb{Z}_2$ insulators.
We would leave the investigation of observational consequences of these phenomena for the future works.

Our analyses so far are based on free theories.
It is interesting to see how the interaction effects change the story.
In interacting systems, 
while the ground state cannot be decomposed 
as a product the Slater determinants as Eq.\ (\ref{gs for non-interacting}),
the Berry connection and the Bures metric can be defined, in principle, 
by twisting the boundary conditions \cite{NiuThoulessWu}.

A yet another possible approach for interacting systems 
is to note that the spectral projector 
$Q({\bf k})$ is nothing but the Fourier transform of the 
equal time correlation function
$\langle \Phi| \psi^{\dag}_i (\mathbf{r})\psi^{\ }_j(0)
|\Phi \rangle$
\cite{Peschel}.
A proper generalization of the momentum space metric 
in interacting systems can then be defined in terms of 
the equal time correlation function.

As an example, let us consider
the (2+1)-dimensional Gross-Neveu model.
I.e., 
the model with the Dirac kinetic term as in (\ref{QHE Dirac Hamiltonian}),
and with a four-fermion interaction. 
To the leading order in the large-$N$ expansion
(here, $N$ is the number of fermion flavors), 
the Euclidean fermion propagate is evaluated as
(see, for example, Ref.\ \cite{Hands1993})
\begin{eqnarray}
\mathcal{G}(k_{\mu})
\!\!&=&\!\!
\frac{ {i}\gamma_{\mu} k^{\mu} -m}{k^2+m^2}
\left(
\frac{\Lambda}{k}
\right)^{a}
\quad
\end{eqnarray}
where
$\mu=0,1,2,3$, 
$k_{\mu=0,1,2,3}$ are the space-time momenta,
$\Lambda$ is the ultra-violet cutoff,
and 
$a=\mathcal{O}(1/N)$. 

The equal time correlator can be read off from 
$\mathcal{G}$ by
$\int^{+\infty}_{-\infty}dk_0  \mathcal{G}(k_0, k_{i})
=
G(k_i)$. 
With the 
equal time correlator,
we can compute the large-$N$ 
corrected quantum metric. 
 The calculation amounts to modifying 
 the projector $Q$ by
$
 Q({\bf k})= \mathcal{H}({\bf k})/\lambda({\bf k})
$
$\to$
$
Q({\bf k}) = \mathcal{H}({\bf k})/\tilde{\lambda}({\bf k}),
$
 where
$ \lambda({\bf k})
 =
\left(k^2 + m^2\right)^{{1\over2}}$,
and
$
 \tilde{\lambda}({\bf k})
 =
 \left(k^2 + m^2\right)^{{1+\gamma\over2}}$. 
 The exponent $\gamma$ is $\mathcal{O}(1/N)$. 
%
The metric is then given in the radial coordinate by
\begin{eqnarray}
ds^2
\!\!&=&\!\! 
\frac{2^n}{8}
\left\{
\left[
\frac{1}{\lambda^{2+2\gamma}}
+
\frac{(\gamma^2-1)k^2}{\lambda^{4+2\gamma}}
\right]
dk^2
+
\frac{k^2 d\Omega_{d-1}}{\lambda^{2+2\gamma}}
\right\}.
\nonumber \\
&&
\end{eqnarray}
In the limit $m\to 0$, this becomes
\begin{eqnarray}
ds^2
= 
\frac{2^n}{8}
\left[
\frac{\gamma^2}{k^{2+2\gamma}}
dk^2
+
\frac{1}{k^{2\gamma}}
d\Omega_{d-1}
\right].
\end{eqnarray}
The momentum space 
is singular at $k=0$ as in the case of non-interacting systems.
However, the structure of the singularity is different.
It would also be interesting to notice that in the absence of interaction, the metric is not 
deformed as we change the number of flavors $N$.
On the other hand, in the interacting case, the metric is deformed as a function of $N$.
This would also lead to deformation in the second cumulant.

Another way of considering the interacting systems is to look at
string theory construction of the topological phases.
As pointed out in \cite{Horava:2005jt} and constructed explicitly in \cite{Ryu:2010hc},
K-theory charges for the topological insulators (superconductors) are closely related
to those of D-branes in string theory.
Since D-brane systems are interacting ones by its construction, it would be interesting to
consider momentum space geometries from D-brane point of view.



\begin{acknowledgments}
We acknowledge 
 ``Quantum Criticality and the AdS/CFT correspondence'' 
 miniprogram at KITP.
 We would like to thank J.\ Polchinski and H.\ Katsura for useful discussion.
The research of SM was supported in part by the National Science Foundation under Grant No. PHY05-51164 and JSPS.
SR thanks the Center for Condensed Matter Theory at University of California,
Berkeley for its support.

\end{acknowledgments}


\end{document}